\documentclass[preprint,showpacs,preprintnumbers,amsmath,amssymb,superscriptaddress]{revtex4}

\usepackage{graphicx}
\usepackage{dcolumn}
\usepackage{color}
\usepackage{ulem}

\begin{document}

\preprint{APS/123-QED}

\title{Many-body two-quantum coherences in semiconductor nanostructures}

\author{Denis Karaiskaj}
\altaffiliation[Present address: ]{Department of Physics, University of South Florida, Tampa, FL 33620}
\affiliation{JILA, National Institute of Standards and Technology and
 University of Colorado, Boulder, CO 80309-0440, USA}

\author{Alan D. Bristow}
\affiliation{JILA, National Institute of Standards and Technology and
 University of Colorado, Boulder, CO 80309-0440, USA}

\author{Lijun Yang}
\affiliation{Chemistry Department, University of California, Irvine,
 California, 92697-2025, USA}%

\author{Xingcan Dai}
\affiliation{JILA, National Institute of Standards and Technology and
 University of Colorado, Boulder, CO 80309-0440, USA}

\author{Richard P. Mirin}
\affiliation{National Institute of Standards and Technology, Boulder, CO 80305, USA}%

\author{Shaul Mukamel}
\affiliation{Chemistry Department, University of California, Irvine,
 California, 92697-2025, USA}

\author{Steven T. Cundiff}
\email{cundiffs@jila.colorado.edu} \affiliation{JILA, National Institute of
Standards and Technology and
 University of Colorado, Boulder, CO 80309-0440, USA}

\date{\today}

\begin{abstract}
We present experimental coherent two-dimensional Fourier transform spectra of the Wannier exciton resonances in semiconductor quantum wells generated by a pulse sequence that isolates two-quantum coherences. By measuring the real part of the signals, we can determine that the spectra are dominated by two-quantum coherences due to mean-field many-body interactions, rather than bound biexcitons. Simulations performed using dynamics controlled truncation agree well with the experiments.

\end{abstract}

\maketitle

The optical properties of semiconductor nanostructures at low temperature are dominated by Wannier excitonic resonances close to the fundamental gap \cite{2007Meier_Book}. Extensive past studies have shown that the nonlinear optical response of the excitonic resonances is dominated by many-body interactions \cite{2001Chemla_Nature}. Most of these studies have used various forms of coherent optical spectroscopy (see reference \onlinecite{2008Cundiff_OpEx} and references therein). Recently, greater insight has been obtained by extending the methods of multidimensional Fourier transform spectroscopy, originally developed in NMR \cite{1987Ernst_Book}, into the optical domain (for a recent review, see, for example, reference \onlinecite{2009Abramavicius_ChemRev}). In semiconductors, two-dimensional Fourier transform (2DFT) spectra have provided insight into the microscopic details of the many-body interactions \cite{2006Li_PRL,2007Zhang_PNAS} showing that many-body correlation terms beyond a mean-field approximation dominate.

One of the hallmarks of many-body interactions has been the appearance of a signal for negative delay in two-pulse transient four-wave-mixing (TFWM) experiments. For an ensemble of non-interacting two-level systems, there is no
signal for negative delay \cite{1979Yajima_JPSJ}. These signals can be
phenomenologically described as arising from local field effects
\cite{1990Leo_PRL,1990Wegener_PRA}, excitation induced dephasing
\cite{1993Wang_PRL,1994Hu_PRB}, biexcitonic effects \cite{1993Bott_PRB} or excitation induced shifts \cite{2002Shacklette_PRB}. While the early work was motivated by explaining the negative delay signal, these effects also contribute, indeed dominate, for positive delay. TFWM could not reliably distinguish between these phenomena, an ambiguity resolved by 2DFT spectroscopy \cite{2006Li_PRL,2007Zhang_PNAS}. The 2D extension of this technique for molecular vibrations was proposed \cite{1999Zhang_JCP,2002Tortschanoff_JCP} and observed \cite{2004Fulmer_JCP,2006Sul_JPCB}. Its sensitivity to two exciton
correlations in semiconductors was demonstrated in simulations
\cite{2008Yang_PRL,2008Yang_PRB}.

In this letter, we present 2DFT spectra for the pulse sequence that isolates two-quantum coherences \cite{2004Fulmer_JCP,2006Sul_JPCB} and corresponds to the ``negative'' delay case in a two-pulse TFWM experiment. In TFWM, the sample is excited by two pulses, $E_1(t+\tau)$ and $E_2(t)$ with wavevectors $\mathbf{k}_1$ and $\mathbf{k}_2$, respectively. Their interaction produces a signal $E_s \propto E_1^*E_2E_2$ in the direction $\mathbf{k}_s = 2 \mathbf{k}_2 - \mathbf{k}_1$ where the delay, $\tau$, between the excitation pulses is defined to be positive if the conjugated pulse, $E_1^*$, arrives before pulse $E_2$. Thus ``negative'' delay means that the conjugated field arrives last. Theory has shown that 2DFT spectra for this pulse ordering are very sensitive to two-exciton correlations \cite{2008Yang_PRL,2008Yang_PRB}. For excitonic resonances in semiconductors, we observe two-quantum coherences due four-particle interactions between two electrons and two holes, which we call ``unbound two-excitons'', in addition to those due to bound biexcitons, which were expected \cite{1996Ferrio_PRB,2009Stone_Science}. By measuring the real part of the two-quantum coherences, as opposed to the magnitude as was done previously \cite{2009Stone_Science}, we are able to resolve the different contributions and confirm that the dominant one is due to many-body interactions between one exciton and the mean-field. Specific polarization of the excitation fields suppresses the biexciton contribution and isolates the remaining many-body effects. Our findings are based on comparison to theoretical simulations.

2DFT spectroscopy is based on a 3-pulse TFWM experiment where the excitation pulses have wave vectors $\mathbf{k}_a$, $\mathbf{k}_b$ and $\mathbf{k}_c$ and the signal is emitted in the direction $\mathbf{k}_s = - \mathbf{k}_a + \mathbf{k}_b + \mathbf{k}_c$. The subscripts, $a, b$ and $c$ do not designate time ordering. The positive delay between first two pulses is designated as $\tau$, between the second and third pulse is $T$ and the signal is emitted during time $t$. The complex signal field, i.e., with full phase information, is measured by heterodyne detection and the delays between the excitation pulses are controlled with interferometric precision, which are not done in TFWM. The time domain multidimensional signal is designated $S_i(\tau,T,t)$ where $i$ specifies the time ordering of the conjugated pulse, $E_a^*$. To generate a 2DFT spectrum, a Fourier transform is taken with respect to two of the time variables while the third is held constant. The ability to take a Fourier transform is based on having measured the complex signal field and the precise stepping of the excitation delays. The frequency domain representation reveals coupling between resonances as off-diagonal (cross) peaks because the absorbing and emitting frequencies will differ \cite{1987Ernst_Book}. Two-quantum coherences appear in $S_{III}(\tau,\omega_T,\omega_t)$ spectra, where $i = III$ designates that the conjugated pulse, $E_a^*$, arrives third.

Measurement of $S_{III}(\tau,\omega_T,\omega_t)$ requires phase stabilization of all excitation pulses, which was not done in previous reports of 2DFT spectra of semiconductors \cite{2006Li_PRL,2007Zhang_PNAS,2005Zhang_OpEx}, but has recently been developed \cite{2009Bristow_Manuscript}. To properly decompose the 2DFT spectra into their real and imaginary parts, phase offsets in the measurement must be corrected. Previously this was done for $S_{I}(\omega_{\tau},T,\omega_t)$ spectra, where $E_a^*$ is the first pulse, by comparison to an auxiliary experiment. An appropriate auxiliary experiment does not exist for $S_{III}(\tau,\omega_T,\omega_t)$, thus we used a recently developed method based on the spatial interference patterns of the excitation beams \cite{2008Bristow_OpEx}.

Figure 1(a) shows a schematic diagram of the experimental setup, which uses the ``box" geometry to generate the TFWM signal. The TFWM signal is heterodyne detected with the reference beam. The time delay $T$ between pulses $\mathbf{k}_c$ and $\mathbf{k}_a$ is scanned whereas the time delay $\tau$ between pulses $\mathbf{k}_b$ and $\mathbf{k}_c$ is kept constant. The Fourier transform of the heterodyne spectra leads to a 2D plot as shown in Fig.~2, where the energy of $\omega_{T}$ is twice that of $\omega_{t}$, indicating that two-quantum transitions are being probed. The sample consists of four periods of a 10 nm GaAs quantum well separated by Al$_{0.3}$Ga$_{0.7}$As barriers grown by molecular beam epitaxy. The sample is held at 6 K.

\begin{figure}
\includegraphics[width=0.5\textwidth]{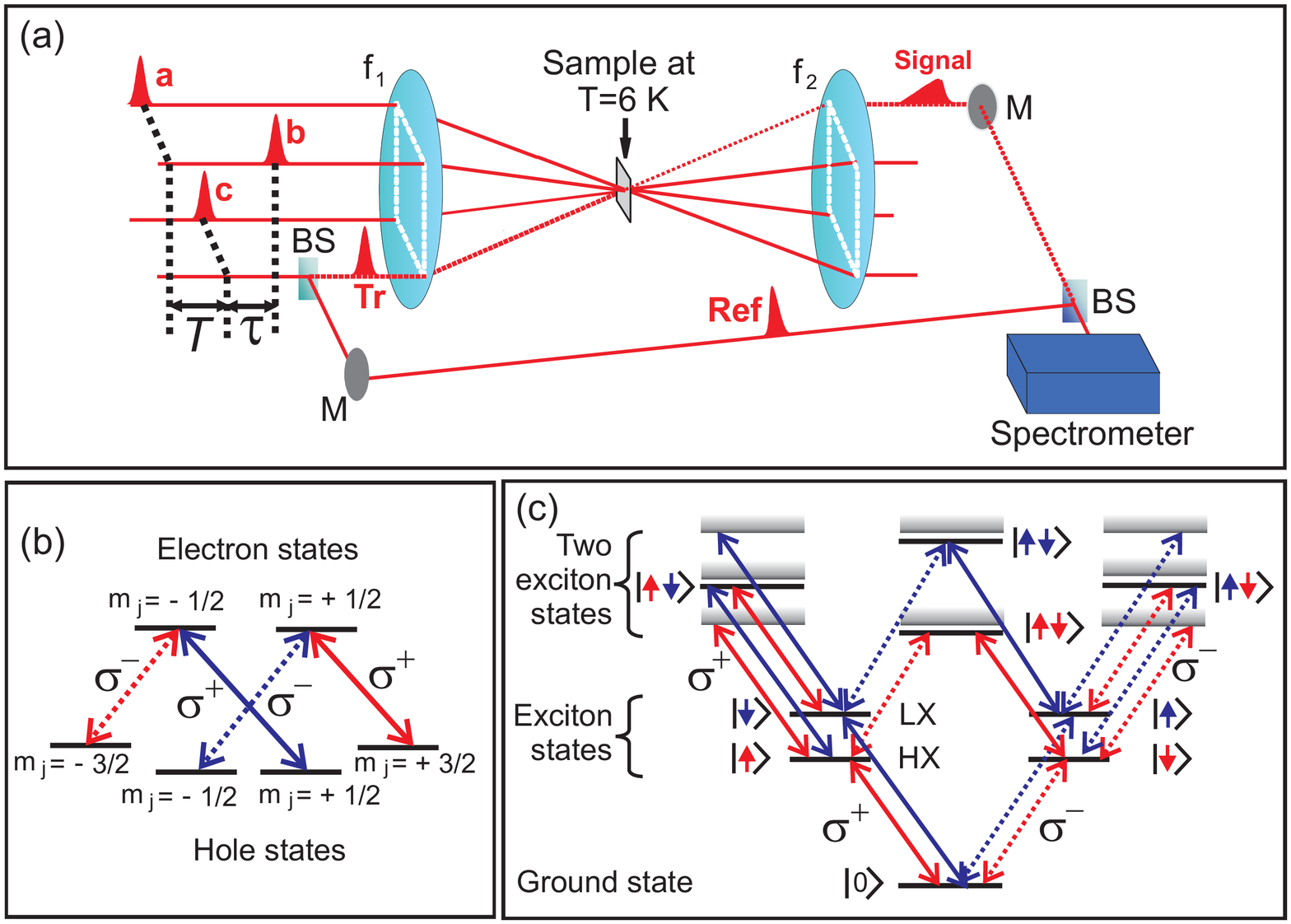}
\caption{\label{fig:epsart} (Color online) (a) Schematic diagram of the experimental setup for optical 2DFT spectroscopy. The incoming laser pulses are focused in the sample by the lens \emph{f$_{1}$}. The box is reconstructed by the lens \emph{f$_{2}$} and the TFWM signal is heterodyne detected with the reference (Ref) beam. (b) Level scheme for the heavy- and light-hole states in GaAs quantum wells. The dashed (solid) arrows represent $\sigma^{-}$ ($\sigma^{+}$) circular polarized light. (c) Level scheme for the heavy- and light-hole excitons and biexcitons in GaAs quantum wells. The 2-exciton states include bound (biexciton) states shown as lines and unbound states indicated by the shaded areas.}
\end{figure}

\begin{figure}
\includegraphics[width=0.5\textwidth]{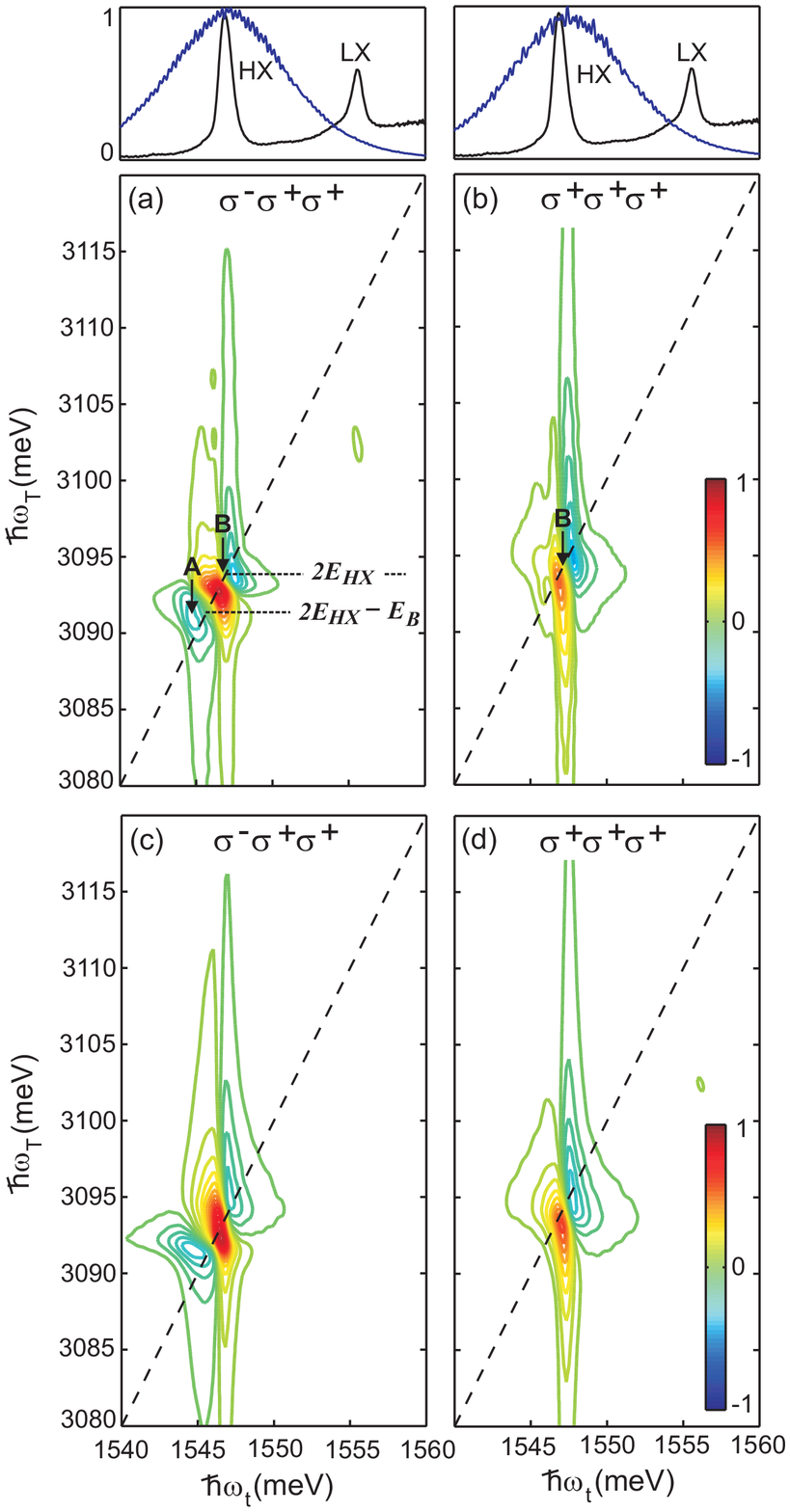}
\caption{\label{fig:epsart} (Color) 2D two-quantum spectra of GaAs quantum wells at different polarizations. (a) Experimental spectra obtained using cross-circular ($\sigma^{-}$$\sigma^{+}$$\sigma^{+}$) polarization and (b) using co-circular ($\sigma^{+}$$\sigma^{+}$$\sigma^{+}$) polarization. The laser
excitation shown above with the absorbance spectra is resonant with HX energy, therefore only the HX peak can be observed. $E_{HX}$ is the heavy-hole exciton energy, and $E_{B}$ is the biexciton binding energy. (c) and (d) show simulations using the same polarizations.}
\end{figure}

The electronic single particle states in the minimum of the conduction band in bulk GaAs are $s$-type two-fold degenerate states with spin sublevels $m_{j}=\pm\frac{1}{2}$, whereas the hole states in the maximum of the valence band are four-fold degenerate with spin sublevels of $m_{j}=\pm\frac{1}{2}$ and $m_{j}=\pm\frac{3}{2}$. Quantum confinement in the $z$-direction in quantum wells lowers the symmetry, lifting the orbital degeneracy of the valence band but preserving the spin degeneracy, and leading to two bands labeled based on their effective mass as heavy hole, with $m_{j}=\pm\frac{3}{2}$, and light hole with $m_{j}=\pm\frac{1}{2}$. Circularly polarized light couples spin states as shown in Fig.~1(b). Excitons can form between holes in either valence band and electrons in the conduction band. Furthermore, excitons with opposite spin electrons can form a bound four particle molecular state known as a biexciton. Biexcitons are not easily discussed in the single particle picture of Fig. 1(b). It is better to use many-particle excitonic states, as shown in Fig.~1(c), which also allow for easier description of two-quantum coherences. Possible transitions are shown in Fig.~1(c) where the excitonic states are labeled by their hole states. In addition to bound biexcitonic states, denoted by a line, in Fig.~1(c) we indicate unbound 2-exciton states by a shaded region. Often unbound 2-exciton states are ignored, however our result will show that they result in the strongest 2-quantum coherences. In Fig.~1(c) the first excitation leads to exciton formation, where HX (LX) refers to an exciton formed by an electron and a heavy (light) hole. A second excitation can lead to a two-quantum coherence. Proceeding along the left side of the ``V" shaped diagram in Fig.~1(c) corresponds to co-circular ($\sigma^{+}$$\sigma^{+}$$\sigma^{+}$)
excitation. In this case a two quantum coherence is formed between the ground state and either mixed bound biexcitons, consisting of a HX and LX, or unbound two-excitons. Cross-circular ($\sigma^{+}$$\sigma^{-}$$\sigma^{-}$) excitation accesses pure HX or LX two-quantum coherences, both bound biexcitons and unbound two-excitons.

Figure 2 shows experimental and theoretical 2DFT
$S_{III}(\tau,\omega_T,\omega_t)$ spectra for two polarization configurations. For reference, the linear absorption spectrum of the sample and the spectrum of the exciting pulses are shown above the 2DFT spectra. Peaks corresponding the HX and LX are evident in the linear spectrum. The laser has been tuned to the HX, thus it dominates the 2DFT spectra shown in Fig.~2(a-d).

Previous theory had predicted that two two-quantum coherences involving bound biexcitons should appear as separate peaks in the 2D spectra for each of the HX and LX transitions \cite{2008Yang_PRL,2008Yang_PRB}. For the heavy-hole exciton, these peaks occur at energies ($\omega_{t} = E_{HX}$, $\omega_T = 2E_{HX}-E_{B}$) and ($E_{HX}-E_{B}$, $2E_{HX}-E_{B}$) where $E_{HX}$ is the heavy-hole exciton energy and $E_{B}$ the biexciton binding energy. In addition, theory shows peaks due to unbound two-exciton  states at the bare two exciton energy, ($E_{HX}$, $2E_{HX}$), as well as scattering states at higher energies ($E_{HX}$, $2E_{HX}+E_{B}$) and ($E_{HX}+E_{B}$, $2E_{HX}+E_{B}$).  The presence and strength of these peaks strongly depended on the polarization configuration.

In the experimental 2D spectra for cross-circular polarization shown in Fig.~2(a), we can identify two peaks, an absorptive peak at energies
($E_{HX}-E_{B}$, $2E_{HX}-E_{B}$) labeled ``A'', and a dispersive peak at energies ($E_{X}$, $2E_{X}$) labeled ``B''. No scattering peaks are evident at $\hbar \omega_T = 2 E_HX+E_B$. However, these scattering peaks only appeared in simulations when the excitation pulse spectrum was adjusted to enhance them \cite{2008Yang_PRL,2008Yang_PRB}. Based on previous theoretical results, the bound biexciton peak A is expected to dominate the cross-circular spectra. In the weak-field coherent limit, cross-circular polarization excitation should produce a small contribution from unbound two-excitons at ($E_{x}$, $2E_{x}$). Surprisingly, peak B dominates the experimental results.  For co-circular polarization, shown in Fig.~2(b), the bound biexciton peak vanishes, as expected. The dominance of peak B in both polarization configurations indicates  that the mean-field contributions dominate over the higher order correlations. The dispersive line shape of peak B also indicates the many-body origin of this transition
\cite{2006Li_PRL,2007Zhang_PNAS}, differing from the absorptive line shape of the bound biexciton peak A.

The simulations presented in Fig.~2 used a simple three-band one-dimensional tight-binding model \cite{1994aAxt_ZP,2007Meier_Book,2008Yang_PRL,2008Yang_PRB}.
The kinetic part of the Hamiltonian is \begin{equation}
\mathbf{H_{kin}}=\sum_{ijc}T_{ij}^{c}c_{i}^{c\dagger}c_{j}^{c}+\sum_{ijv}T_{ij}^{v}%
d_{i}^{v\dagger}d_{j}^{v},\label{0}%
\end{equation}
where $c_{i}^{c\dagger}(c_{i}^{c})$ are creation (annihilation) Fermi operators of electrons in site $i$ from the conduction band $c$ and
$d_{i}^{v\dagger}(d_{i}^{v})$ are the corresponding operators for holes in the valence band $v$. The diagonal elements $T_{ii}^{c,v}$ describe the site energies for the electrons (holes) in the conduction (valence) band while off-diagonal elements $T_{i\neq j}^{c,v}$ represent the couplings between different sites. The dipole interaction with the radiation field has the form $\mathbf{H_{int}}= -E \cdot P$, where $P$ is the interband polarization operator and $E$ is the electric field. By adding a Coulomb interaction potential, we obtain the total Hamiltonian
\begin{equation}
\mathbf{H}=\mathbf{H_{kin}}+\mathbf{H_{int}}+\mathbf{H_{C}}.
\end{equation}
With the exception of the biexciton binding energy, which is determined by an interplay between the inter-site coupling and Coulomb interaction, all parameters are determined based on the known effective masses and the experimental absorption spectrum. The present analysis is constrained to the coherent-limit, where only single-exciton and two-exciton variables are relevant. The Heisenberg equations of motion of these two variables are derived from the total Hamiltonian and are truncated according to the leading order in the external field. The biexciton binding energy and dephasing rates are adjusted to get good agreement with the experimental 2DFT spectra. The 2D signals are then calculated by selecting the spatial Fourier
components of the equations of motion. Although this theoretical model is relatively simple it successfully reproduces basic features of the experimental spectra. In particular, it reproduces (1) the dominance of peak B under both polarization conditions, (2) the behavior of peak A under both polarization conditions, and (3) the line shapes of both peaks. Namely, absorptive for peak A and dispersive for peak B.

\begin{figure}
\includegraphics[width=0.45\textwidth]{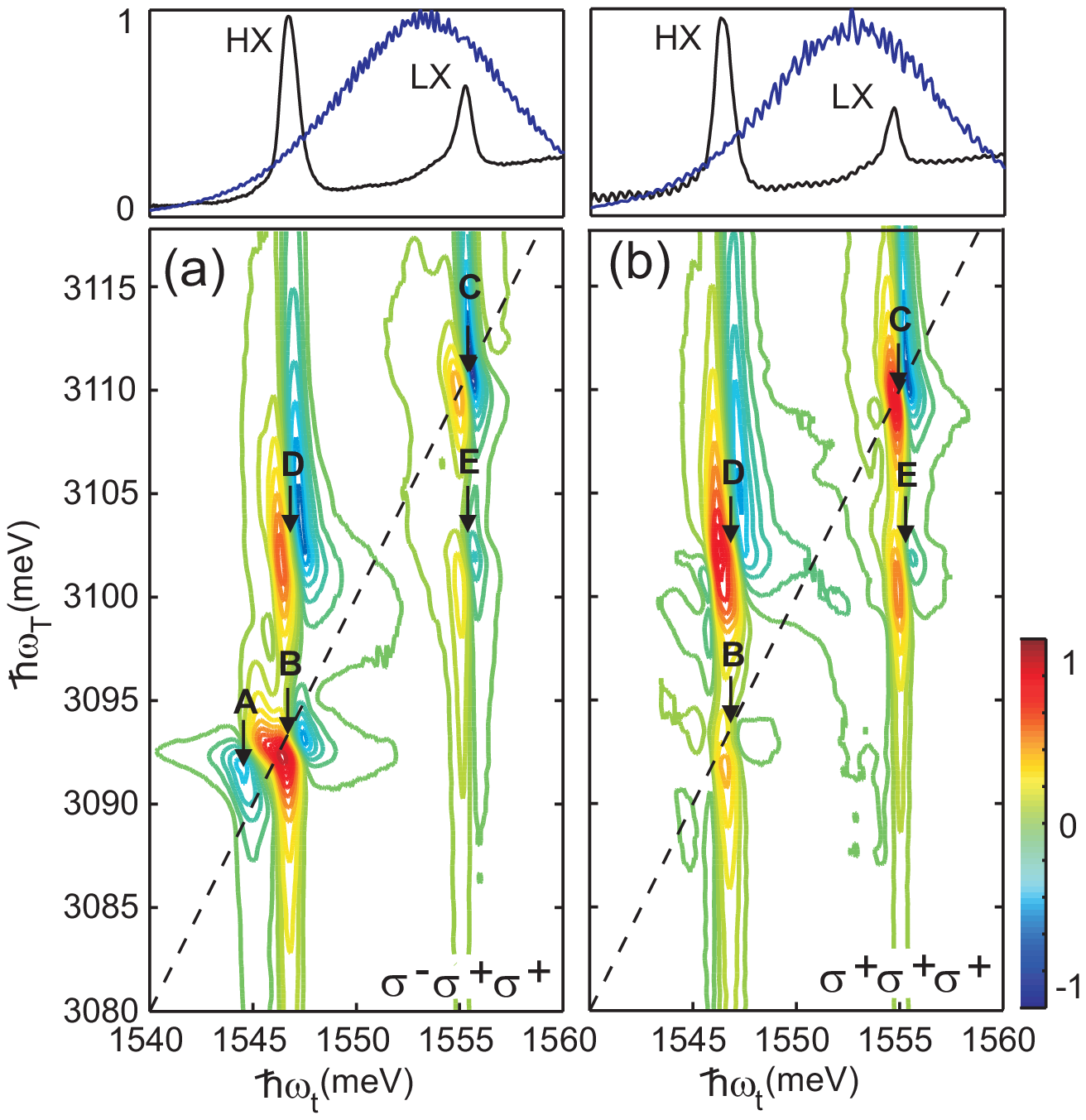}
\caption{\label{fig:epsart} (Color) Experimental 2D two-quantum spectra of GaAs quantum wells when both excitons are excited. (a) Spectrum obtained using cross-circular ($\sigma^{-}$$\sigma^{+}$$\sigma^{+}$)
polarization and (b) using co-circular ($\sigma^{+}$$\sigma^{+}$$\sigma^{+}$)
polarization. Laser excitation shown above with the absorption spectrum.}
\end{figure}

To observe two-quantum transitions originating from HX and LX, we tuned the laser excitation spectrum between the HX and LX excitation energies. The excitation and absorption spectra are depicted at the top of Fig.~3. The two HX peaks labeled as A and B in Fig.~2 can be observed in Fig.~3(a) as well, whereas the LX shows only one peak (C) at ($E_{LX}$, $2E_{LX}$). It is not surprising that the LX, due to its lower effective mass, does not form bound states but is entirely dominated by the unbound two-exciton peak C \cite{2009Bristow_PRB}. In
addition, two cross peaks above the diagonal (D) and below the diagonal (E) are observed, at the bare exciton energies $E_{HX}$ or $E_{LX}$, respectively, along $\omega_{t}$. The presence and strength of these peaks indicates strong two-quantum coherent coupling between the HX and LX states. As a result of the dominance of the unbound two-exciton peaks B and C for both polarization conditions, the cross peaks D and E are present for both polarizations (Fig.~3(a) and (b)).

To gain insight into the physical origin of the observed spectral peaks, we repeated the calculations neglecting higher order correlations beyond mean-field. We find that the peaks corresponding to bound biexcitons disappear, as expected. The unbound two-exciton peaks remain, with little change in strength or lineshape. This result leads us to conclude that the mean-field terms dominate these two-quantum spectra. While higher-order correlations are needed to match the experiment, they do not dominate to the same extent found in previous one-quantum spectra \cite{2007Zhang_PNAS}.

In conclusion, we observe many-body signatures in the real part of the 2DFT $S_{III}$ spectra of semiconductor quantum wells. The 2D spectra collected using co-circular and cross-circular polarizations reveal unexpected results. Co-circular polarization spectra are dominated by the unbound many-body two-exciton peak at the HH energy, as predicted theoretically. Surprisingly, the 2D spectra for cross-circular polarization are dominated by the many-body two-exciton peak at the HH energy, although a bound biexciton peak emerges. Strong cross peaks between HH and LH two-exciton transitions indicate significant many-body two-quantum coherent coupling. A deeper understanding of these many-body effects may lead the development of coherent processes in semiconductors that would not be predicted from a simple level scheme. For example, electromagnetically induced transparency (EIT) due to exciton correlations has been observed in semiconductors \cite{2002Phillips_PRL}. Most likely, this implementation of EIT was exploiting the same unbound two-exciton states that we have discovered.

This work was supported by  the Chemical Sciences, Geosciences, and Biosciences
Division Office of Basic Energy Sciences, (U.S.) Department of Energy and the
National Science Foundation.

\end{document}